\documentstyle[sprocl,epsfig]{article}

\def\be{\begin{equation}}
\def\ee{\end{equation}}
\def\bea{\begin{eqnarray}}
\def\eea{\end{eqnarray}}

\begin{document}

\begin{flushright}
{\large UG-FT-164/04} 

{\large CAFPE-34/04}
\end{flushright}

\vspace{0.1cm} 

\title{FORMULAE FOR A NUMERICAL COMPUTATION OF ONE-LOOP TENSOR INTEGRALS
\footnote{Talk given at LCWS2004, Paris, France, April 2004.\\
Work supported by the EU under contract HPRN-CT-2002-00149 and by MECD under contract SAB2002-0207.}
}

\author{ROBERTO PITTAU
\footnote{On leave of absence from Dipartimento di 
Fisica Teorica, Torino and INFN Sezione di Torino, Italy.}
}
\address{Departamento de F{\'\i}sica Te\'orica y del Cosmos 
and Centro Andaluz de 
F{\'\i}sica de Part{\'\i}culas Elementales (CAFPE), 
Universidad de Granada, E-18071 Granada, Spain
}


\maketitle\abstracts{A numerical and iterative approach for computing 
one-loop tensor integrals is presented.}

\section{Introduction}
In Reference \cite{delAguila:2004nf} a new approach has been introduced for computing,  
recursively and numerically, one-loop tensor integrals. Here we describe a few modifications 
of the original method that allow a more efficient numerical implementation 
of the algorithm. We keep all of our notations as in Ref. \cite{delAguila:2004nf} and, in particular,
put a bar over $n$-dimensional quantities and a tilde over $\epsilon$-dimensional objects
($n= 4+\epsilon$). 
The formulae we want to modify are Eqs. (9), (35), (48), (52) and (54) of Ref. 
\cite{delAguila:2004nf}, 
that, all together, allow to reduce any $(m+1)$-point tensor integral with $m\,\ge \,2$ 
to the standard set of scalar one-loop functions \cite{'tHooft:1978xw}.

Such formulae are not symmetric when interchanging any pair of loop denominators $\bar D_k$, 
because are derived under the assumption that at least 
one of them (identified with $\bar D_0$) carries a vanishing external momentum, namely  
\bea
\label{eq:1}
\bar D_k= (\bar q+p_k)^2-m^2_k\,,~~k=0,\cdots ,m\,,~~p^\mu_0= 0 \,.
\eea
Already after the first iteration, terms appear in which the denominator $\bar D_0$ is
canceled by a $\bar D_0$ reconstructed in the numerator, so that the resulting integrals do not fulfill 
any longer the assumption of Eq. (\ref{eq:1}). A shift of the integration variable $\bar q$ 
is then needed to bring them back to a form suitable to apply the algorithm again.
However, shifting $\bar q$ may generate a large amount of terms, 
especially when dealing with high rank tensors, so that 
deriving more symmetric formulae, in which $p^\mu_0 \ne 0$, is clearly preferable. 

A second useful modification is related to the problem outlined in Sec. 6 of Ref. 
\cite{delAguila:2004nf}, that occur 
when $p_1^2 = 0$ and $p_2^2 \ne 0$ ($p_1^2 \ne 0$ and $p_2^2= 0$) 
and $(p_1 \cdot p_2) \sim  0$.
For those kinematical configurations a new linear combination of the momenta $p_1$ and $p_2$
is needed as a basis of the reduction to ensure the numerical stability of
the algorithm. Once again, it is better to include such cases right from the beginning.
\section{The General Recursion Formula}
When $p^\mu_0 \ne 0$, the $n$-dimensional version of Eq. (9) of Ref. \cite{delAguila:2004nf} 
should be modified as follows
\bea
\label{eq:2}
&&\!\!\!\!\!\!\!\!\!\!I^{(n)}_{m;\,\mu \nu \rho \cdots \tau}  ~=~ \frac{\beta}{2 \gamma}\, 
T_{\mu  \nu \lambda \sigma}\,\left\{J^{(n)\,\lambda\sigma}_{m;\,\rho \cdots \tau}
\right\}   \\
&-&\frac{1}{4 \,\gamma} T_{\mu \nu} 
\left\{(m_0^2-p_0^2)\, I^{(n)}_{m;\,\rho \cdots \tau} 
+I^{(n)}_{m-1;\,\rho \cdots \tau}(0)-2\,p_{0\alpha} I^{(n)\,\alpha}_{m;\,\rho \cdots \tau} 
-I^{(n;\,2)}_{m;\,\rho \cdots \tau}
\right\}  \nonumber \\
&-&\frac{1}{4 \,\gamma}
T_{\mu \nu \lambda} 
\left\{h_{3} I^{(n)\,\lambda}_{m;\,\rho \cdots \tau}      
             +I^{(n)\,\lambda}_{m-1;\,\rho \cdots \tau}(3)
             -I^{(n)\,\lambda}_{m-1;\,\rho \cdots \tau}(0)
 -\frac{2 \beta}{\gamma} k_{3\alpha}\,J^{(n)\,\alpha\lambda}_{m;\,\rho \cdots \tau}
 \right\}\,, \nonumber
\eea
where
\bea
\label{eq:3}
J^{(n)\,\lambda\sigma}_{m;\,\rho \cdots \tau}\!\! &=&\!\!
(h_{1} r_2^\lambda + h_{2} r_1^\lambda)\, 
I^{(n)\,\sigma}_{m;\,\rho \cdots \tau} 
+(r_2^\lambda + \xi_2 r_1^\lambda)\, I^{(n)\,\sigma}_{m-1;\,\rho \cdots \tau}(1)  \\
\!\!&+& \!\! (r_1^\lambda + \xi_1 r_2^\lambda)\, I^{(n)\,\sigma}_{m-1;\,\rho \cdots \tau}(2)
-[r_1^\lambda (1+\xi_2)+r_2^\lambda (1+\xi_1)]\, I^{(n)\,\sigma}_{m-1;\,\rho \cdots \tau}(0)\,.
\nonumber
\eea
and where the extra integrals $I^{(n;\,2)}_{m;\,\rho \cdots \tau}$ are defined in Eq. (77) of 
Ref.  \cite{delAguila:2004nf}.

In the previous Equations, $k_i= p_i-p_0$ and the massless 4-momenta $\ell_{1,2}$
to be used as a basis of the reduction algorithm, as in Eq. (13) of Ref.  \cite{delAguila:2004nf}, 
are such that
\be
\label{eq:4}
s_1 = \ell_1+\alpha_1 \ell_2\,,~~~s_2 =  \ell_2+\alpha_2 \ell_1\,,
\ee
where $s_{1,2}$ are suitable linear combinations of $k_{1,2}$  
\be
s_1 = k_1+\xi_1 k_2\,,~~~s_2 =  k_2+\xi_2 k_1\,.
\ee
By choosing
\be
\xi_1= \frac{1}{2}\,\,{\rm sign}(k_2^2)\,\,{\rm sign}(k_1 \cdot k_2)\,~~{\rm and}~~  
\xi_2= \frac{1}{2}\,\,{\rm sign}(k_1^2)\,\,{\rm sign}(k_1 \cdot k_2)\,,
\ee
the quantity
\be
\gamma= \frac{s_1^2 s_2^2 }{(s_1 \cdot s_2) \mp \sqrt{\Delta}} \equiv (s_1 \cdot s_2) \pm \sqrt{\Delta}
\,,~~~ \Delta =  (s_1 \cdot s_2)^2-s_1^2 s_2^2\,, 
\ee
defined in Eq. (62) of Ref. \cite{delAguila:2004nf} only vanishes
when $k_1^2 = k_2^2 = (k_1 \cdot k_2) = 0$, that always corresponds to collinear configurations 
cut away in physical observables, therefore solving the second problem outlined in the Introduction.
The tensors $T_{\mu  \nu \lambda \sigma}\,, T_{\mu  \nu \lambda},\, T_{\mu  \nu}$
and the 4-vectors $r_{12}$ are defined as in
Ref. \cite{delAguila:2004nf}, but in terms of $\ell_{1,2}$ given in Eq. (\ref{eq:4}) and 
with the replacement $p_3 \to k_3$.
Finally
\bea
h_1 &=& (m_1^2-p_1^2) + \xi_1\,(m_2^2-p_2^2)-(1+\xi_1)\,(m_0^2-p_0^2)\,, \nonumber \\
h_2 &=& (m_2^2-p_2^2) + \xi_2\,(m_1^2-p_1^2)-(1+\xi_2)\,(m_0^2-p_0^2)\,, \nonumber \\
h_3 &=& (m_3^2-p_3^2) - (m_0^2-p_0^2)\,, \nonumber \\
\frac{\beta}{\gamma}  &=&  \pm \frac{1}{2\,\sqrt{\Delta}}\,.
\eea
The derivation of Eq. (\ref{eq:2}) closely follows the derivation of Eq. (9) 
of  Ref. \cite{delAguila:2004nf}. For example, choosing $\ell_{1,2}$ as 
in Eq. (\ref{eq:4}), the quantity 
\bea
D_{\mu} = \frac{1}{\beta}\, [2\,(q\cdot\ell_1) \ell_{2\mu}+2\,(q\cdot\ell_2) \ell_{1\mu}]
\eea
defined in Eq. (18) of Ref. \cite{delAguila:2004nf} can be rewritten as
\bea 
D_\mu &=& [\bar D_1+ \xi_1 \bar D_2 - (1+\xi_1) \bar D_0 + h_1]\,r_{2\mu} \nonumber \\
      &+& [\bar D_2+ \xi_2 \bar D_1 - (1+\xi_2) \bar D_0 + h_2]\,r_{1\mu}\,,
\eea
and generates the term $J^{(n)}$ of Eq. (\ref{eq:3}).
Analogously, choosing $b= k_3$ in Eq. (22) of Ref. \cite{delAguila:2004nf}, generates the first
part of the last term of Eq. (\ref{eq:2}), because, when $p_0^\mu \ne 0$
\bea
2\,(q \cdot k_3) = \bar D_3-\bar D_0 + h_3\,.
\eea
Finally, the equality
\bea
q^2= \bar D_0+(m_0^2-p_0^2)-2\,(q\cdot p_0)-{\tilde q}^2\,,
\eea
is the origin of the second row of Eq. (\ref{eq:2}).

\section{Three-point Tensors}
When $p^\mu_0 \ne 0$, rank 2 and rank 3 three-point tensor integrals need a separate
treatment. The relevant formulae follow by adapting the theorems in  
Eqs. (37) and (40) of Ref. \cite{delAguila:2004nf} to the case  $p^\mu_0 \ne 0$:
\bea
\label{eq:theorem}
&&\int d^n \bar q \frac{1}{\bar D_0\bar D_1\bar D_2} 
[(q+p_0) \cdot \ell_{3}]^i= 0\,, \nonumber \\
&&\int d^n \bar q \frac{1}{\bar D_0\bar D_1\bar D_2} [(q+p_0) \cdot \ell_4]^i= 0\,,
~~~~~~\forall\, i=\,1,2,3\cdots \,~~{\rm and} \nonumber \\
&&\int d^n \bar q \frac{1}{\bar D_0\bar D_1\bar D_2} 
 [(q+p_0) \cdot \ell_{3,4}]^2\,q_\rho 
~=~ 0\,. 
\eea
The final results read as follows
\bea
\label{eq:5}
I^{(n)}_{2;\,\mu \nu}  &=& \frac{\beta}{2 \gamma}\, 
T^\prime_{\mu  \nu \lambda \sigma}\,\left\{J^{(n)\,\lambda\sigma}_{2}
\right\}
~-~\frac{1}{4 \,\gamma} t_{\mu \nu} 
\left\{(m_0^2-p_0^2)\,  I^{(n)}_{2} 
+I^{(n)}_{1}(0) \right. \nonumber \\
&-& \left. 2\,p_{0 \alpha} I^{(n)\,\alpha}_{2} 
-I^{(n;\,2)}_{2} \right\}\,
~-~\frac{1}{4 \,\gamma} T^{\prime}_{\mu \nu \lambda} 
\left\{p_0^{\lambda}\,I^{(n)}_{2} \right\}\,, \nonumber \\
I^{(n)}_{2;\,\mu \nu \rho}  &=& \frac{\beta}{2 \gamma}\, 
T^\prime_{\mu  \nu \lambda \sigma}\,\left\{J^{(n)\,\lambda\sigma}_{2;\,\rho}
\right\}
~-~\frac{1}{4 \,\gamma} t_{\mu \nu} 
\left\{(m_0^2-p_0^2)\,  I^{(n)}_{2;\,\rho} 
+I^{(n)}_{1;\,\rho}(0) \right. \nonumber \\
&-& \left. 2\,p_{0 \alpha} I^{(n)\,\alpha}_{2;\,\rho} 
-I^{(n;\,2)}_{2;\,\rho} \right\}\,
~-~\frac{1}{4 \,\gamma} T^{\prime}_{\mu \nu \lambda} 
\left\{-p_0^{\lambda}\,I^{(n)}_{2;\,\rho}-2\,I^{(n)\,\lambda}_{2;\,\rho}
 \right\}.
\eea
$J^{(n)}$ is given in Eq. (\ref{eq:3}) and
\bea
\label{eq:6}
t_{\mu \nu} &=& \ell_{3\mu}\ell_{4\nu}+\ell_{4\mu}\ell_{3\nu}\,, \nonumber \\
T^{\prime}_{\mu \nu \lambda} &=& - 
\frac{\ell_{3\mu}\ell_{3\nu}\ell_{4\lambda}(p_0 \cdot \ell_4)+
      \ell_{4\mu}\ell_{4\nu}\ell_{3\lambda}(p_0 \cdot \ell_3)}{\gamma}\,. 
\eea
\section{Rank One Tensors}
In this section we adapt Eqs. (48), (52) and (54) of Ref. \cite{delAguila:2004nf}
to the case $p^\mu_0 \ne 0$.
\subsection{The $m=2$ case}
With $t_{\alpha \mu}$ defined in Eq. (\ref{eq:6}) we get
\bea
I^{(n)}_{2;\,\mu}= \frac{\beta}{\gamma} J^{(n)}_{2;\,\mu} + \frac{p_0^\alpha}{2\gamma}
\,t_{\alpha \mu}\,I^{(n)}_{2}\,. 
\eea
\subsection{The $m=3$ case}
With $T_{\mu \nu \lambda}$ defined as in Eq. (23) of Ref. \cite{delAguila:2004nf} we get
\bea
I^{(n)}_{3;\,\mu}  &=& \frac{\beta}{\gamma}\,J^{(n)}_{3;\,\mu} 
~+~\frac{1}{4}\,\left[
\frac{\ell_{3\mu}}{(k_3 \cdot \ell_3)}+
\frac{\ell_{4\mu}}{(k_3 \cdot \ell_4)}
\right] \nonumber \\ 
&\times& 
\left\{h_{3} I^{(n)}_{3}      
             +I^{(n)}_{2}(3)
             -I^{(n)}_{2}(0)
 -\frac{2 \beta}{\gamma} k_{3}^{\lambda}\,J^{(n)}_{3;\,\lambda} \right\} \nonumber \\
 &-&\frac{1}{4 \gamma}\,T_{\mu \nu \lambda}\,(p_0^\nu k_3^\lambda-p_0^\lambda k_3^\nu)\,I^{(n)}_{3}\,.
\eea
\subsection{The $m>3$ case}
The generalization of Eq. (54) of  Ref. \cite{delAguila:2004nf} reads
\bea
I^{(n)}_{m;\,\mu}  &=& \frac{\beta}{\gamma}
J^{(n)}_{m;\,\mu}
+ \frac{\ell_{3\mu}\ell_{4\alpha}-\ell_{3\alpha}\ell_{4\mu}}{2\,\delta} \nonumber \\
&\times&\left\{
 k_3^\alpha
\left[h_{4}  I^{(n)}_{m}      
             +I^{(n)}_{m-1}(4)
             -I^{(n)}_{m-1}(0)
 -\frac{2 \beta}{\gamma} k_{4\lambda}\,J^{(n)\,\lambda}_{m}
 \right] \right. \nonumber \\
&-& k_4^\alpha \left.
\left[h_{3}  I^{(n)}_{m}      
             +I^{(n)}_{m-1}(3)
             -I^{(n)}_{m-1}(0)
 -\frac{2 \beta}{\gamma} k_{3\lambda}\,J^{(n)\,\lambda}_{m}
 \right]
\right\}\,,
\eea
where $\delta = (\ell_3\cdot k_4)(\ell_4\cdot k_3)-(\ell_3\cdot k_3)(\ell_4\cdot k_4)$, and
$h_4 = (m_4^2-p_4^2) - (m_0^2-p_0^2)$.
\section{The Extra Integrals}
In most practical cases, the extra integrals, such as
$I^{(n;\,2)}_{m;\,\rho \cdots \tau}$ in Eq. (\ref{eq:2}), are either zero or scaleless,
so that, even when $p_0^\mu \ne 0$, one can directly use the results given in 
Appendix B of Ref. \cite{delAguila:2004nf}.
In all other cases modifications are needed. For example, 
Eqs. (78) and (83) of Ref. \cite{delAguila:2004nf} must be replaced by
\bea
I^{(n;\,2)}_{2;\,\mu} &=& \frac{i \pi^2}{6} (p_{0\mu}+p_{1\mu}+p_{2\mu}) 
               + {\cal O}(\epsilon)\,, \nonumber \\ 
I^{(n;\,2)}_{1} &=&  -i \frac{\pi^2}{2}\left[ m_1^2+m_0^2-\frac{(p_1-p_0)^2}{3}\right] 
+ {\cal O}(\epsilon)\,.
\eea
\section{Conclusion}
We have derived a set of formulae to efficiently implement the 
$n$-dimensional reduction algorithm presented in Ref. \cite{delAguila:2004nf}.
As for the three-point tensors, we limited our analysis to ranks $\le 3$. For higher ranks, a general 
formula can be easily derived, with the help of Appendix C of Ref. \cite{delAguila:2004nf},
using the theorems of Eq. (\ref{eq:theorem}).

\end{document}